\documentclass[preprint, showkeys, groupaddress, superscriptaddress]{revtex4-1}

\usepackage{amssymb}
\usepackage{amsthm}
\usepackage{textcomp}
\usepackage{graphicx}
\usepackage{hyperref}

\begin{document}

\title{Magnetic skyrmions in confined geometries : effect of the magnetic field and the disorder}

\author{Rom\'eo Juge}
\affiliation{SPINTEC, CEA-INAC, CNRS, Universit\'e Grenoble Alpes, Grenoble INP, F-38000 Grenoble, France.}

\author{Soong-Geun Je}
\affiliation{SPINTEC, CEA-INAC, CNRS, Universit\'e Grenoble Alpes, Grenoble INP, F-38000 Grenoble, France.}

\author{Dayane de Souza Chaves}
\affiliation{Institut N\'eel, CNRS, Universit\'e Grenoble Alpes, 25 avenue des Martyrs, B.P. 166, Grenoble Cedex 9 38042, France.}

\author{Stefania Pizzini}
\affiliation{Institut N\'eel, CNRS, Universit\'e Grenoble Alpes, 25 avenue des Martyrs, B.P. 166, Grenoble Cedex 9 38042, France.}

\author{Liliana D. Buda-Prejbeanu}
\affiliation{SPINTEC, CEA-INAC, CNRS, Universit\'e Grenoble Alpes, Grenoble INP, F-38000 Grenoble, France.}

\author{Lucia Aballe}
\affiliation{ALBA Synchrotron Light Facility, Carrer de la Llum 2-26, 08290 Cerdanyola del Vall\`es (Barcelona), Spain.}

\author{Michael Foerster}
\affiliation{ALBA Synchrotron Light Facility, Carrer de la Llum 2-26, 08290 Cerdanyola del Vall\`es (Barcelona), Spain.}

\author{Andrea Locatelli}
\affiliation{Elettra Sincrotrone, S.C.p.A, S.S 14 - km 163.5 in AREA Science Park 34149 Basovizza, Trieste, Italy.}

\author{Tevfik Onur Mente\c{s}}
\affiliation{Elettra Sincrotrone, S.C.p.A, S.S 14 - km 163.5 in AREA Science Park 34149 Basovizza, Trieste, Italy.}

\author{Alessandro Sala}
\affiliation{Elettra Sincrotrone, S.C.p.A, S.S 14 - km 163.5 in AREA Science Park 34149 Basovizza, Trieste, Italy.}

\author{Francesco Maccherozzi}
\affiliation{Diamond Light Source, Chilton, Didcot, Oxfordshire, OX11 0DE, UK.}

\author{Sarnjeet S. Dhesi}
\affiliation{Diamond Light Source, Chilton, Didcot, Oxfordshire, OX11 0DE, UK.}

\author{St\'ephane Auffret}
\affiliation{SPINTEC, CEA-INAC, CNRS, Universit\'e Grenoble Alpes, Grenoble INP, F-38000 Grenoble, France.}

\author{Gilles Gaudin}
\affiliation{SPINTEC, CEA-INAC, CNRS, Universit\'e Grenoble Alpes, Grenoble INP, F-38000 Grenoble, France.}

\author{Jan Vogel}
\affiliation{Institut N\'eel, CNRS, Universit\'e Grenoble Alpes, 25 avenue des Martyrs, B.P. 166, Grenoble Cedex 9 38042, France.}

\author{Olivier Boulle}
\email{olivier.boulle@cea.fr}
\affiliation{SPINTEC, CEA-INAC, CNRS, Universit\'e Grenoble Alpes, Grenoble INP, F-38000 Grenoble, France.}

\begin{abstract}

We report on the effect of the lateral confinement and a perpendicular magnetic field on isolated room-temperature magnetic skyrmions in sputtered Pt/Co/MgO nanotracks and nanodots. We show that the skyrmions size can be easily tuned by playing on the lateral dimensions of the nanostructures and by using external magnetic field amplitudes of a few mT, which allow to reach sub-100 nm diameters. Our XMCD-PEEM observations also highlight the important role of the pinning on the skyrmions size and stability under an out-of-plane magnetic field. Micromagnetic simulations reveal that the effect of local pinning can be well accounted for by considering the thin film grain structure with local anisotropy variations and reproduce well the dependence of the skyrmion diameter on the magnetic field and the geometry.

\end{abstract}

\maketitle

\section{Introduction}

Magnetic skyrmions are nanoscale whirling spin configurations \cite{nagaosa_topological_2013}. Their small size, topological protection and the fact that they can be manipulated by small in-plane current densities have opened a new paradigm to manipulate the magnetization at the nanoscale. This has led to proposals for novel memory and logic devices in which the magnetic skyrmions are the information carriers \cite{fert_skyrmions_2013}. They were first observed in B20 chiral magnets thin films \cite{muhlbauer_skyrmion_2009, yu_real-space_2010} within confined geometries \cite{du_edge-mediated_2015, jin_control_2017}, and in ultrathin epitaxial films \cite{heinze_spontaneous_2011, romming_writing_2013, romming_field-dependent_2015} at low temperature. The recent observation of room-temperature magnetic skyrmions \cite{buttner_dynamics_2015, jiang_blowing_2015, woo_observation_2016-1, boulle_room-temperature_2016-1, moreau-luchaire_additive_2016-1, soumyanarayanan_tunable_2016, legrand_room-temperature_2017, pollard_observation_2017, schott_skyrmion_2017, hrabec_current-induced_2017} and their current-induced manipulation \cite{jiang_blowing_2015, woo_observation_2016-1, legrand_room-temperature_2017, yu_room-temperature_2016, litzius_skyrmion_2017, woo_spin-orbit_2017, hrabec_current-induced_2017} in sputtered magnetic films have lifted an important bottleneck toward the practical realization of such devices. However, these experiments have also underlined the sensitivity of the skyrmions dynamics to the defects in the materials as well as the edges which can impede reliable motion. This raises the question of stability - nucleation and annihilation - in these sputtered ultrathin magnetic films. More recently, the influence of the granular structure and the inherent disorder of these films on the skyrmions stability have begun to be addressed \cite{legrand_room-temperature_2017, hrabec_current-induced_2017, kim_current-driven_2017-1}.

Here we report on the observation of the effect of the lateral confinement, a perpendicular magnetic field and the local pinning sites on the magnetic
texture in sputtered Pt/Co/MgO ultrathin nanostructures with isolated room-temperature magnetic skyrmions. In particular, the evolution of the mean skyrmion diameter in sub-micrometers nanotracks and nanodots is followed by high resolution magnetic imaging. These experimental results are supported by micromagnetic simulations which reproduce well the magnetic field dependence of the skyrmion size in the different geometries. Notably, we show that the granular structure of the sputtered materials leads to local pinning of the skyrmion which can strongly affect its shape and size. Our results underline that the local pinning and the sample's geometry play an important role in the field dependence of the skyrmion size and its stability in ultrathin films.

\section{Methods}

We used X-ray Magnetic Circular Dichroism combined with Photo-Emission Electron Microscopy (XMCD-PEEM) at room-temperature to obtain a direct image of the magnetization pattern with a high lateral spatial resolution (down to $\sim$ 30 nm). The observed contrast is proportional to the projection of the magnetization on the incident X-ray beam direction, which is impinging on the sample under a grazing angle of 16{\textdegree}. As a result, the magnetic contrast is about 3.5 times larger for the magnetization in the sample plane as compared to the magnetization perpendicular to it. As shown in our previous work \cite{boulle_room-temperature_2016-1}, this allows the direct observation of the internal spin structure of magnetic domain walls (DWs) or magnetic skyrmions. The XMCD-PEEM experiments were carried out with the SPELEEM III microscope (Elmitec GmbH) at the Nanospectroscopy beamline at the Elettra synchrotron in Basovizza, Trieste, Italy, at the CIRCE beamline at the Alba synchrotron, Barcelona, Spain \cite{aballe_alba_2015, foerster_custom_2016} and at the IO6 beamline at the Diamond synchrotron, Didcot, UK.

The Ta(3)/Pt(3)/Co(0.5-1.1)/MgO$_x$/Ta(2) (thickness in nm) film was deposited by magnetron sputtering on a 100 mm high-resistivity Si wafer, then annealed for 1.5 h at 250 {\textdegree}C under vacuum and an in-plane magnetic field of 240 mT. The Co layer was deposited as a wedge. The magnetic domains in the thin film are  perpendicularly magnetized. All the images presented correspond to a nominal Co thickness of about 1.0 nm, an area close to the reorientation transition and were obtained at room temperature.

The micromagnetic simulations were performed using the Mumax3 code \cite{vansteenkiste_design_2014}. We used the following micromagnetic parameters extracted from our previous experiments \cite{boulle_room-temperature_2016-1} : the Co layer thickness t = 1.06 nm, the exchange constant A = 2.75$\times$10$^{-11}$ J.m$^{-1}$, the uniaxial anisotropy constant K$_u$ = 1.45$\times$10$^{6}$ J.m$^{-3}$, the interfacial Dzyaloshinskii-Moriya interaction constant D = 2.05$\times$10$^{-3}$ J.m$^{-2}$, and the saturation magnetization M$_s$ = 1.44$\times$10$^{6}$ A.m$^{-1}$. For all the geometries considered hereafter, the cells size is 1 nm $\times$ 1 nm $\times$ 1.06 nm with only one cell across the film thickness, so that the magnetization is supposed to be uniform along this direction. For convenience, all the simulations were performed at T = 0 K, although temperature should play a significant role on the skyrmion stabilization and interaction with the local disorder \cite{rohart_path_2016}. Finally, we define $d_s$ as the mean skyrmion diameter, namely the diameter that would have a perfectly circular bubble with the same area as the distorted one.

\section{Results and discussion}

\subsection{Observation of room-temperature magnetic skyrmions in nanotracks}

\begin{figure}
\includegraphics[scale=0.79]{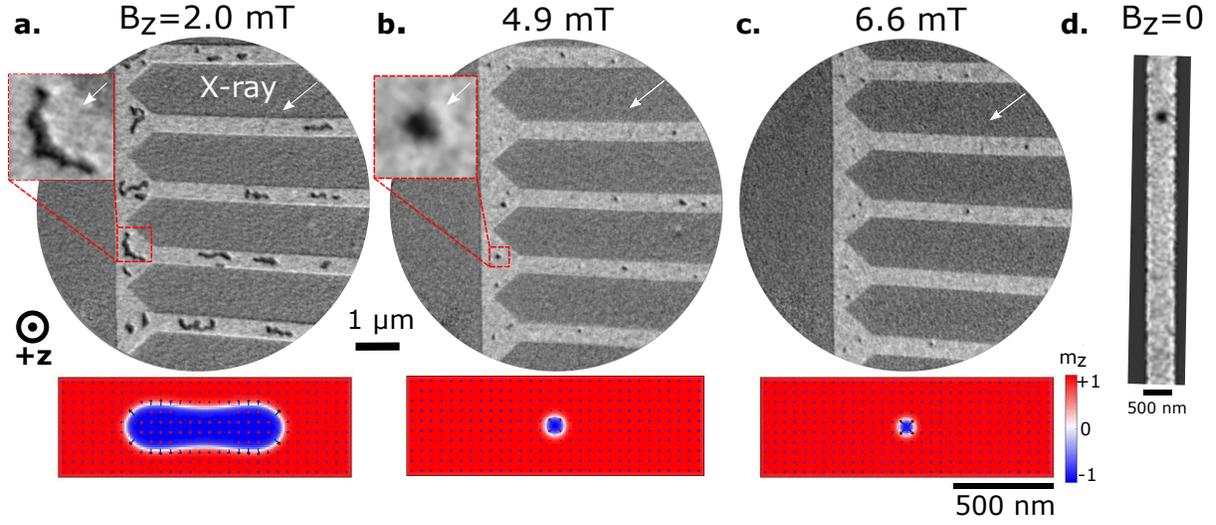}
\centering
\caption{\textbf{a-c.} XMCD-PEEM images of the magnetic texture in 500 nm-wide tracks during the application of a magnetic field perpendicular to the film plane. The insets show a zoom in the region delimited by the dashed red boxes of dimensions \textbf{a.} 800 nm $\times$ 800 nm and \textbf{b.} 500 nm $\times$ 500 nm. Below each image is represented the simulated distribution of the magnetization in 1500 nm $\times$ 500 nm tracks, for the same applied magnetic field and in the case of a disorder-free film. The white arrows indicate the X-ray beam direction. \textbf{d.} XMCD-PEEM image of a magnetic skyrmion in a 540 nm-wide track at zero external magnetic field.}
\label{tracks images}
\end{figure}

At remanence or in the presence of a small magnetic field ($<$ 2 mT), worm-like domains are observed in the tracks (Fig.\ref{tracks images}.a), a configuration which minimizes the dipolar energy. These worms tend to align with the tracks, while they are randomly oriented in the largest areas of the nanostructures. Moreover, they seem to be repelled from the edges, which is reproduced well by the simulations. Interestingly, a strong white (black) contrast is observed on the upper (bottom) DWs delimiting the worm-like domains (see inset of Fig.\ref{tracks images}.a). As the contrast is proportional to the projection of the magnetization along the X-ray propagation direction, and the DWs are perpendicular to this direction, the contrast is consistent with DWs having a left-handed homochiral N\'eel structure \cite{boulle_room-temperature_2016-1}. This internal DW structure is reproduced well by the micromagnetic simulations. When applying a larger magnetic field (typically B$_z>$ 2 mT) the initial worm-like domains shrink and lead to the formation of isolated skyrmions (Fig.\ref{tracks images}.b-c). Again, the white/black contrast at the top/bottom edge of the skyrmion demonstrates that the homochiral N\'eel structure is conserved (see inset of Fig.\ref{tracks images}.b). \\

\begin{figure}
\includegraphics[scale=1]{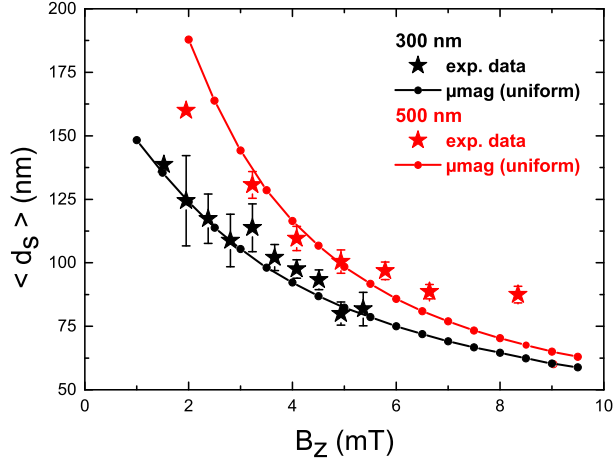}
\centering
\caption{Average skyrmion diameter $<d_s>$ as a function of the applied magnetic field measured in 500 nm-wide tracks (red stars) and 300 nm-wide tracks (black stars). The error bars represent the standard deviation. The red dots (black dots) correspond to the micromagnetic simulations in the case of a 500 nm-wide track (300 nm-wide track).}
\label{tracks graph}
\end{figure}

The average skyrmion diameter is plotted as a function of $B_z$ in Fig.\ref{tracks graph} for tracks of two different widths (300 nm and 500 nm) along with the results of the simulations. For both wire widths, a large decrease of the skyrmions size is observed when small fields ($\sim$ 5 mT) are applied. In the 300 nm tracks, the average diameter changes from 140 nm down to 80 nm, showing a large susceptibility of the skyrmions size to B$_z$. In addition, we observe that the track width affects the skyrmions size in particular at low magnetic field : the narrower the track the smaller the skyrmions, as suggested by previous numerical studies \cite{zhang_skyrmion-skyrmion_2015-1}. This geometrical confinement effect can be explained by the dipolar interactions of the DWs delimiting the skyrmion with the magnetic charges on the tracks edges. These results thus show that the skyrmion size can be easily tuned by playing both on the lateral confinement and the magnetic field \footnote{A similar field dependence of the skyrmions size has been observed in larger structures, as reported by S. Woo \textit{et al.} in 2 {\textmu}m-diameter Pt/Co/Ta disks (see \cite{woo_observation_2016-1}, Supplementary Information). Nevertheless, in our case, the skyrmions number and their size are limited by the track width at small fields.}. As the magnetic field increases and the skyrmion diameter decreases, the influence of the track width on the skyrmion becomes smaller, indicating that the skyrmion equilibrium size is governed by the Zeeman energy term and that the dipolar interaction with the track edges plays a less important role. However, we also see experimentally that, for B$_z=$ 6.7 mT, some of the skyrmions are not visible anymore and even more have disappeared at B$_z=$ 8.5 mT. In fact they are too small to be observed, since by decreasing the magnetic field down to $\sim$ 1 mT (not shown), the skyrmions expand and turn into worm-like domains at the same position as they were initially. Interestingly, we observed in 540 nm-wide tracks, that magnetic skyrmions can be stabilized also at zero external magnetic field (see Fig.\ref{tracks images}.d) ; this can be attributed to the presence of local pinning sites. \\

In the following section, we present the results of the observation of magnetic skyrmions in a Pt/Co/MgO circular nanodot and we discuss the role played by the local disorder on the field dependence of the skyrmion size.

\subsection{Magnetic skyrmion in a circular nanodot}

\begin{figure}
\includegraphics[scale=0.86]{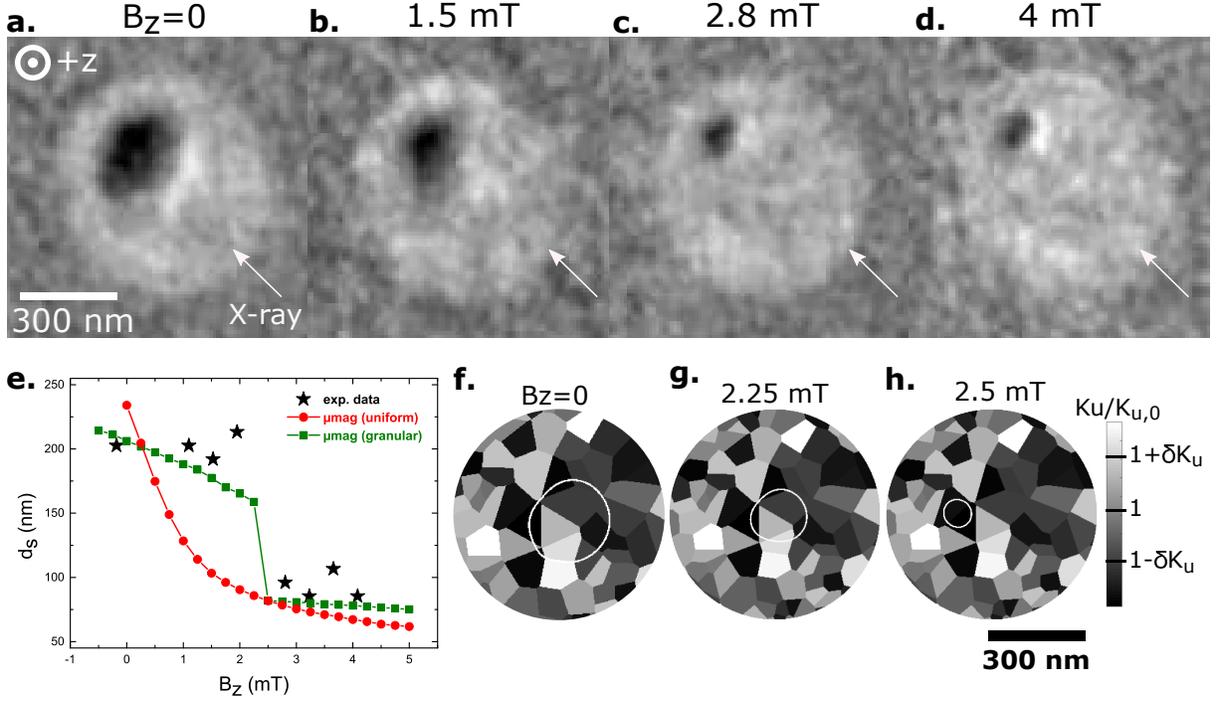}
\centering
\caption{\textbf{a-d.} XMCD-PEEM images of a magnetic skyrmion in a 630 nm-diameter circular dot for different applied magnetic field amplitudes. \textbf{e.} Skyrmion diameter as a function of the applied magnetic field. The experimental data points (black stars) are extracted from the XMCD-PEEM images. Micromagnetic simulations : the red dots correspond to the case of a perfect defect-free film and the green squares correspond to the case of a disordered film with $g=$ 70 nm and $\delta$K$_u$ = 2.5 \%. \textbf{f-h.} Representation of the skyrmion border (center of the N\'eel wall, m$_z$ = 0) on the grain structure for \textbf{f.} B$_z$ = 0 (initial state relaxed at zero field), \textbf{g.} B$_z$ = 2.25 mT (before the jump) and \textbf{h.} B$_z$ = 2.5 mT (after the jump). m$_z<$ 0 inside the white border and m$_z>$ 0 outside. The color scale indicates the amplitude of the anisotropy fluctuations.}
\label{dot}
\end{figure}

The images presented in Fig.\ref{dot}.a. show a magnetic skyrmion in a 630 nm-diameter circular dot at zero external magnetic field with a diameter of about 190 nm. We first observe that the skyrmion is positionned near the edge of the dot, whereas the dipolar repulsion from the edges favor a skyrmion position in the center of the dot. This suggests that the skyrmion is pinned by local defects near the edge of the dot. The application of a small perpendicular magnetic field does not change substantially the skyrmion structure and its size (Fig.\ref{dot}.b, B$_z$ = 1.5 mT) up to a critical field B$_z$ = 2.8 mT where a sudden decrease of the skyrmion diameter to 90 nm is observed (see Fig.\ref{dot}.c-d and stars in Fig.\ref{dot}.e). Moreover, the skyrmion appears slightly distorted for B$_z<$ 2.4 mT while it is circularly shaped for B$_z>$ 2.4 mT. \\

To better understand these experimental results, we performed micromagnetic simulations in the case of a perfect disorder-free film. The initial state is a 200 nm-diameter circular bubble placed at the center of the 630 nm-diameter circular dot ; this configuration is then relaxed to the minimum energy state at zero field. When the magnetic field increases, the skyrmion diameter decreases continuously and covers a wider range of sizes than the one observed (see Fig.\ref{dot}.e, red dots). Moreover, the contraction of the core is isotropic and the skyrmion stays in the center of the dot (not shown), as expected notably from the symmetry of the demagnetizing field. \\

Is it well known that the polycrystalline grain structure of magnetic ultrathin films is a source of local pinning for magnetic DWs \cite{ranjan_grain_1987, yu_pinning_1999, voto_effects_2016}. To model it, we introduce a grain distribution with a fixed average grain size $g$. From grain to grain, we assume that the uniaxial anisotropy constant fluctuates randomly following a Gaussian distribution with a mean value K$_{u,0}$ (the measured macroscopic value) and a standard deviation $\delta$K$_uK_{u,0}$, where $\delta$K$_u$ quantifies the pinning strength \cite{kim_current-driven_2017-1, voto_effects_2016, garcia-sanchez_skyrmion-based_2016}. The grain size and the anisotropy were varied respectively within the ranges 10 nm $\leq{}g\leq$ 100 nm and 2.5 \% $\leq\delta$K$_u\leq$ 10 \% with multiple realizations for each set of parameters. Our simulations reveal a very similar behavior to the experimental one for several realizations of grains distributions (\textit{i.e.} for given sets of parameters \{g, $\delta$K$_u$\}). We show in Fig.\ref{dot}.e. (green squares) the dependence of the skyrmion diameter on the perpendicular magnetic field for a specific realization, shown in Fig.\ref{dot}.f-h, with $g=$ 70 nm and $\delta$K$_u$ = 2.5 \%. As it is observed experimentally, the skyrmion diameter decreases in a step-like fashion when the magnetic field reaches a certain critical value. On the images presented in Fig.\ref{dot}.f-h, the center of the DW delimiting the skyrmion is represented on the grain structure at zero field, before (B$_z$ = 2.25 mT) and after the jump (B$_z$= 2.5 mT) respectively ; with m$_z<$ 0 inside the white border and m$_z>$ 0 outside. The color scale indicates the amplitude of the anisotropy constant in each grain. The case of relatively large grains ($g=$ 70 nm) has been chosen for clarity but other grain configurations allow to reproduce this step-like behavior. For B$_z\leq$ 2.25 mT, the equilibrium position and shape of the skyrmion is determined by the grains of weakest anisotropy (darkest regions) so that, in order to lower the DW energy, the skyrmion core is stretched away from its initial circular shape and its initial position in the center of the dot. In this field range, the skyrmion is trapped in its initial position by the local grain structure and the magnetic field affects little its shape and position. When the depinning field (B$_z$ = 2.5 mT) is reached, the skyrmion prefers to jump in a neighboring region where the anisotropy is smaller leading to smaller DW energy despite the  additional cost in   dipolar energy away for the dot center. Thus, the polycrystalline grain structure of the sputtered material allows to reproduce the experimentally observed step-like behavior when applying an external magnetic field, as well as the local pinning near the edge.

\subsection{Statistical study : influence of the grain size g and the pinning strength $\delta$K$_u$}

In the previous section, we discussed one particular example and showed that one can find configurations which reproduce well the behavior of the skyrmion in a nanodot under an out-of-plane magnetic field. Here we investigate the influence of the characteristic parameters of the disorder : the grain size $g$ and the pinning strength $\delta$K$_u$. For each set of parameters \{$g$, $\delta$K$_u$\}, the diameter is averaged over multiple realizations and $<d_s>$ will refer to the skyrmion diameter averaged over 50 different realizations ($g$ and $\delta$K$_u$ being fixed). \\

In Fig.\ref{Ku2.5, ds vs g, slope}.a, $<d_s>$ is plotted as a function of the applied magnetic field for grain sizes between 10 nm and 100 nm and for $\delta$K$_u$ = 2.5\%. For comparison, the case of an ideal, disorder-free film (red dots) and the experimental measurements (black stars) are also represented. It appears that, on average, the skyrmion size and its field dependence are little affected by the grain size, except for the smallest one considered ($g=$ 10 nm) where skyrmions are less sensitive to the pinning at high field. In fact, this behavior is observed for larger values of $\delta$K$_u$ (up to 10 \% in this study), meaning that the skyrmion is more sensitive to the applied field for $g=$ 10 nm than for all the other grain sizes considered here. However, for $g$ ranging between 20 nm and 100 nm, there is no monotonic variation of the diameter with the grain size. We note that this seems in contrast with previous numerical reports \cite{kim_current-driven_2017-1, voto_effects_2016} where a maximum in the DW depinning field was observed when the mean grain size equals the DW width $\pi\Delta$. We do not observe such a maximum in our simulations when $g$ = $\pi{}\Delta{}$ = 40 nm. \\

To complete the picture, the graph of Fig.\ref{Ku2.5, ds vs g, slope}.b shows $<d_s>$ as a function of $g$ at zero applied field and for two different pinning strengths : $\delta$K$_u$ = 2.5 \% and $\delta$K$_u$ = 10 \%. We observe that at low pinning strength, the skyrmion diameter is approximately constant with the grain size, whereas it increases on average for larger pinning strength, but this is negligible for $g\leq$ 50 nm. Note that this remains true at higher fields. In fact, at high pinning, the initial state is stretched out after relaxation and this effect is all the more important as the grain size increases and it very often turns skyrmions into worm-like domains. This results in a small shift (along the y-axis) of the curves $<d_s($B$_z)>$ for $\delta$K$_u$ = 10 \% and  $g\geq$ 50 nm (not shown). However, it does not affect the field dependence, as it is discussed hereafter. On the contrary, for relatively weak pinning strengths ($\delta$K$_u$ = 2.5 \% and $\delta$K$_u$ = 5 \%), the skyrmion recovers its initial circular symmetry at high field, as shown in Fig.\ref{dot}.h. When starting from a large magnetic field and decreasing its amplitude, the quasi-circular skyrmion very often expands into an irregularly-shaped domain, a behavior observed in other systems and attributed to structural defects \cite{pulecio_hedgehog_2016}. An additional interesting feature is that  the standard deviation (error bars) increases with the grain size for both pinning strengths, suggesting that for a strong anisotropy disorder and for a large film surface, for instance a one-dimensional track, one will be more likely to observe an important dispersion in terms of sizes and shapes for large grains. \\

We now discuss the influence of the pinning strength on the magnetic field susceptibility of the skyrmion size. We define $\eta$ as the slope (absolute value) obtained from a linear fit of the curves $<d_s($B$_z)>$ within the field range 2 mT $\leq$ B$_z$ $\leq$ 5 mT and at fixed g and $\delta$K$_u$, consistently with our experimental results (Fig.\ref{tracks graph}). Since, for a given pinning strength, no significant change in this slope with the mean grain size is observed (see Fig.\ref{Ku2.5, ds vs g, slope}.a for the case $\delta$K$_u=$ 2.5 \%), we plot the average value $<\eta>_g$ of the slope $\eta$ over the grain size $g$ as a function of $\delta$K$_u$ in Fig \ref{Ku2.5, ds vs g, slope}.c. We see that $<\eta{}>_g$, which quantifies the sensitivity of the skyrmion to the applied magnetic field, decreases rapidly for $\delta$K$_u$ between 2.5 \% and 7.5 \% and then seems to saturate. When the disorder is too important, in this case $\delta$K$_u\geq$ 7.5 \%, the skyrmion shape and size is fully determined by the grains and the anisotropy mapping and the  small external magnetic field has little influence on the final state. This is also confirmed by the small standard deviation observed for $\delta$K$_u=$ 10 \% : whatever the grain size, when the anisotropy disorder is strong, the effect of the magnetic field becomes negligible. Actually, the probability for a step-like decrease of the diameter falls to zero within this field range as soon as $\delta$K$_u\geq$ 7.5 \%. The field sensitivity of the skyrmion diameter thus provides an indirect way to assess the disorder in the thin film and our experimental value of $\eta=$ 17 nm.mT$^{-1}$ in the 500 nm-wide tracks indicates a disorder $\delta$K$_u$ typically lower than 5 \%.

\newpage

\section{Conclusion}

In conclusion, we have observed that the skyrmion size in Pt/Co(1 nm)/MgO can be easily tuned by reducing the dimensions of the nanostructures and that magnetic field amplitudes of a few mT allow to reach sub-100 nm diameters. These observations also highlight the important role of the pinning on the skyrmion size and stability under an out-of-plane magnetic field. Micromagnetic simulations reveal that the effect of local pinning can be well accounted for by considering the thin film grain structure with local anisotropy variations and reproduce well the dependence of the skyrmion diameter on the magnetic field and the geometry. Our simulation suggests a relatively small anisotropy disorder in our thin films (typically, $\delta$K$_u\leq$ 5 \%). These results also reflect that the Pt/Co/MgO system, due to its simple structure, can be efficiently modeled, which paves the way for further experimental and micromagnetic investigations.

\newpage

\newpage

\newpage

\newpage

\begin{figure}
\includegraphics[scale=0.82]{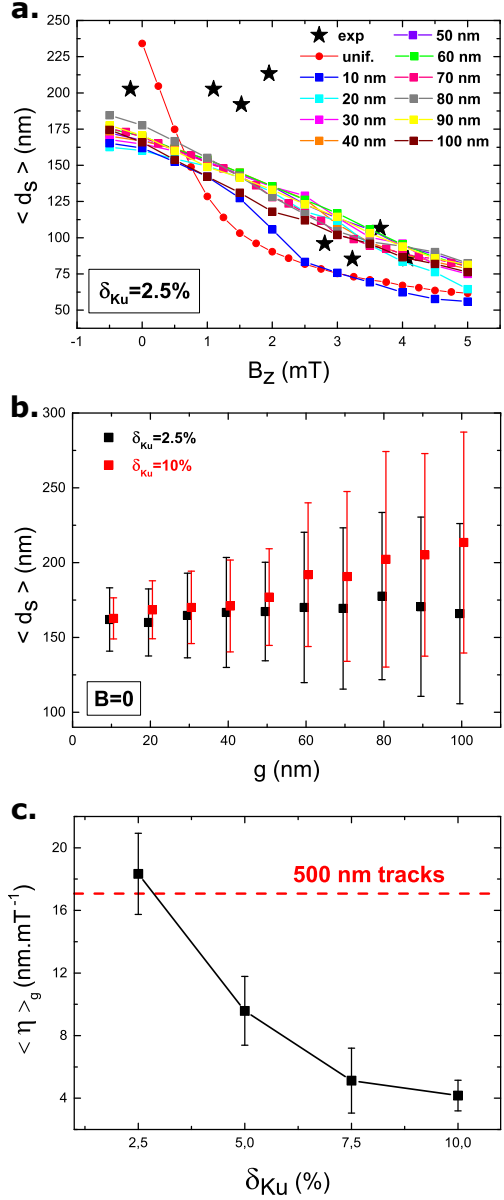}
\centering
\caption{\textbf{a.} Skyrmion diameter averaged over 50 realizations as a function of the applied magnetic field for different grain sizes $g$ from 10 nm to 100 nm in a 630 nm-diameter circular dot and for a fixed pinning strength $\delta$K$_u$ = 2.5 \%. The black stars correpond to the experimental data points and the red stars to the case of a uniform film without disorder. \textbf{b.} Skyrmion diameter averaged over 50 realizations as a function of the grain size at zero applied magnetic field for $\delta$K$_u$ = 2.5 \% (black) and $\delta$K$_u$ = 10 \% (red). The error bars represent the standard deviation. The two curves have been shifted along the x-axis for clarity. \textbf{c.} Slope $\eta$ (absolute value) extracted from a linear fit of the curves $<d_s($B$_z)>$ (at fixed $g$ and $\delta$K$_u$) in the range 2 mT $\leq$ B$_z$ $\leq$ 5 mT and averaged over the grain sizes in the range 10 nm $\leq{}g\leq$ 100 nm. Here, the error bars represent the standard deviation from the average $<\eta{}>_g$ over $g$.}
\label{Ku2.5, ds vs g, slope}
\end{figure}

\clearpage
\newpage

\bibliography{bibliography}

\begin{thebibliography}{10}
\expandafter\ifx\csname url\endcsname\relax
  \def\url#1{\texttt{#1}}\fi
\expandafter\ifx\csname urlprefix\endcsname\relax\def\urlprefix{URL }\fi
\expandafter\ifx\csname href\endcsname\relax
  \def\href#1#2{#2} \def\path#1{#1}\fi

\bibitem{nagaosa_topological_2013}
N.~Nagaosa, Y.~Tokura, Topological properties and dynamics of magnetic
  skyrmions, Nature Nanotechnology 8~(12) (2013) 899--911.
\newblock \href {http://dx.doi.org/10.1038/nnano.2013.243}
  {\path{doi:10.1038/nnano.2013.243}}.

\bibitem{fert_skyrmions_2013}
A.~Fert, V.~Cros, J.~Sampaio, Skyrmions on the track, Nature Nanotechnology
  8~(3) (2013) 152--156.
\newblock \href {http://dx.doi.org/10.1038/nnano.2013.29}
  {\path{doi:10.1038/nnano.2013.29}}.

\bibitem{muhlbauer_skyrmion_2009}
S.~M{\"u}hlbauer, B.~Binz, F.~Jonietz, C.~Pfleiderer, A.~Rosch, A.~Neubauer,
  R.~Georgii, P.~B{\"o}ni, Skyrmion {{Lattice}} in a {{Chiral Magnet}}, Science
  323~(5916) (2009) 915--919.
\newblock \href {http://dx.doi.org/10.1126/science.1166767}
  {\path{doi:10.1126/science.1166767}}.

\bibitem{yu_real-space_2010}
X.~Z. Yu, Y.~Onose, N.~Kanazawa, J.~H. Park, J.~H. Han, Y.~Matsui, N.~Nagaosa,
  Y.~Tokura, Real-space observation of a two-dimensional skyrmion crystal,
  Nature 465~(7300) (2010) 901--904.
\newblock \href {http://dx.doi.org/10.1038/nature09124}
  {\path{doi:10.1038/nature09124}}.

\bibitem{du_edge-mediated_2015}
H.~Du, R.~Che, L.~Kong, X.~Zhao, C.~Jin, C.~Wang, J.~Yang, W.~Ning, R.~Li,
  C.~Jin, X.~Chen, J.~Zang, Y.~Zhang, M.~Tian, Edge-mediated skyrmion chain and
  its collective dynamics in a confined geometry, Nature Communications 6
  (2015) 8504.
\newblock \href {http://dx.doi.org/10.1038/ncomms9504}
  {\path{doi:10.1038/ncomms9504}}.

\bibitem{jin_control_2017}
C.~Jin, Z.-A. Li, A.~Kov{\'a}cs, J.~Caron, F.~Zheng, F.~N. Rybakov, N.~S.
  Kiselev, H.~Du, S.~Bl{\"u}gel, M.~Tian, Y.~Zhang, M.~Farle, R.~E.
  Dunin-Borkowski, Control of morphology and formation of highly geometrically
  confined magnetic skyrmions, Nature Communications 8 (2017) 15569.
\newblock \href {http://dx.doi.org/10.1038/ncomms15569}
  {\path{doi:10.1038/ncomms15569}}.

\bibitem{heinze_spontaneous_2011}
S.~Heinze, K.~{von Bergmann}, M.~Menzel, J.~Brede, A.~Kubetzka,
  R.~Wiesendanger, G.~Bihlmayer, S.~Bl{\"u}gel, Spontaneous atomic-scale
  magnetic skyrmion lattice in two dimensions, Nature Physics 7~(9) (2011)
  713--718.
\newblock \href {http://dx.doi.org/10.1038/nphys2045}
  {\path{doi:10.1038/nphys2045}}.

\bibitem{romming_writing_2013}
N.~Romming, C.~Hanneken, M.~Menzel, J.~E. Bickel, B.~Wolter, K.~von Bergmann,
  A.~Kubetzka, R.~Wiesendanger, Writing and {{Deleting Single Magnetic
  Skyrmions}}, Science 341~(6146) (2013) 636--639.
\newblock \href {http://dx.doi.org/10.1126/science.1240573}
  {\path{doi:10.1126/science.1240573}}.

\bibitem{romming_field-dependent_2015}
N.~Romming, A.~Kubetzka, C.~Hanneken, K.~{von Bergmann}, R.~Wiesendanger,
  Field-{{Dependent Size}} and {{Shape}} of {{Single Magnetic Skyrmions}},
  Physical Review Letters 114~(17) (2015) 177203.
\newblock \href {http://dx.doi.org/10.1103/PhysRevLett.114.177203}
  {\path{doi:10.1103/PhysRevLett.114.177203}}.

\bibitem{buttner_dynamics_2015}
F.~B{\"u}ttner, C.~Moutafis, M.~Schneider, B.~Kr{\"u}ger, C.~M. G{\"u}nther,
  J.~Geilhufe, C.~v.~K. Schmising, J.~Mohanty, B.~Pfau, S.~Schaffert, A.~Bisig,
  M.~Foerster, T.~Schulz, C.~a.~F. Vaz, J.~H. Franken, H.~J.~M. Swagten,
  M.~Kl{\"a}ui, S.~Eisebitt, Dynamics and inertia of skyrmionic spin
  structures, Nature Physics 11~(3) (2015) 225--228.
\newblock \href {http://dx.doi.org/10.1038/nphys3234}
  {\path{doi:10.1038/nphys3234}}.

\bibitem{jiang_blowing_2015}
W.~Jiang, P.~Upadhyaya, W.~Zhang, G.~Yu, M.~B. Jungfleisch, F.~Y. Fradin, J.~E.
  Pearson, Y.~Tserkovnyak, K.~L. Wang, O.~Heinonen, S.~G.~E. te~Velthuis,
  A.~Hoffmann, Blowing magnetic skyrmion bubbles, Science 349~(6245) (2015)
  283--286.
\newblock \href {http://dx.doi.org/10.1126/science.aaa1442}
  {\path{doi:10.1126/science.aaa1442}}.

\bibitem{woo_observation_2016-1}
S.~Woo, K.~Litzius, B.~Kr{\"u}ger, M.-Y. Im, L.~Caretta, K.~Richter, M.~Mann,
  A.~Krone, R.~M. Reeve, M.~Weigand, P.~Agrawal, I.~Lemesh, M.-A. Mawass,
  P.~Fischer, M.~Kl{\"a}ui, G.~S.~D. Beach, Observation of room-temperature
  magnetic skyrmions and their current-driven dynamics in ultrathin metallic
  ferromagnets, Nature Materials 15~(5) (2016) 501--506.
\newblock \href {http://dx.doi.org/10.1038/nmat4593}
  {\path{doi:10.1038/nmat4593}}.

\bibitem{boulle_room-temperature_2016-1}
O.~Boulle, J.~Vogel, H.~Yang, S.~Pizzini, D.~{de Souza Chaves}, A.~Locatelli,
  T.~O. Mente{\c s}, A.~Sala, L.~D. Buda-Prejbeanu, O.~Klein, M.~Belmeguenai,
  Y.~Roussign{\'e}, A.~Stashkevich, S.~M. Ch{\'e}rif, L.~Aballe, M.~Foerster,
  M.~Chshiev, S.~Auffret, I.~M. Miron, G.~Gaudin, Room-temperature chiral
  magnetic skyrmions in ultrathin magnetic nanostructures, Nature
  Nanotechnology 11~(5) (2016) 449--454.
\newblock \href {http://dx.doi.org/10.1038/nnano.2015.315}
  {\path{doi:10.1038/nnano.2015.315}}.

\bibitem{moreau-luchaire_additive_2016-1}
C.~Moreau-Luchaire, C.~Moutafis, N.~Reyren, J.~Sampaio, C.~a.~F. Vaz, N.~V.
  Horne, K.~Bouzehouane, K.~Garcia, C.~Deranlot, P.~Warnicke, P.~Wohlh{\"u}ter,
  J.-M. George, M.~Weigand, J.~Raabe, V.~Cros, A.~Fert, Additive interfacial
  chiral interaction in multilayers for stabilization of small individual
  skyrmions at room temperature, Nature Nanotechnology 11~(5) (2016) 444--448.
\newblock \href {http://dx.doi.org/10.1038/nnano.2015.313}
  {\path{doi:10.1038/nnano.2015.313}}.

\bibitem{soumyanarayanan_tunable_2016}
A.~Soumyanarayanan, M.~Raju, A.~L.~G. Oyarce, A.~K.~C. Tan, M.-Y. Im, A.~P.
  Petrovic, P.~Ho, K.~H. Khoo, M.~Tran, C.~K. Gan, F.~Ernult, C.~Panagopoulos,
  Tunable {{Room Temperature Magnetic Skyrmions}} in {{Ir}}/{{Fe}}/{{Co}}/{{Pt
  Multilayers}}, arXiv:1606.06034 [cond-mat]\href
  {http://arxiv.org/abs/1606.06034} {\path{arXiv:1606.06034}}.

\bibitem{legrand_room-temperature_2017}
W.~Legrand, D.~Maccariello, N.~Reyren, K.~Garcia, C.~Moutafis,
  C.~Moreau-Luchaire, S.~Collin, K.~Bouzehouane, V.~Cros, A.~Fert,
  Room-{{Temperature Current}}-{{Induced Generation}} and {{Motion}} of sub-100
  nm {{Skyrmions}}, Nano Letters 17~(4) (2017) 2703--2712.
\newblock \href {http://dx.doi.org/10.1021/acs.nanolett.7b00649}
  {\path{doi:10.1021/acs.nanolett.7b00649}}.

\bibitem{pollard_observation_2017}
S.~D. Pollard, J.~A. Garlow, J.~Yu, Z.~Wang, Y.~Zhu, H.~Yang, Observation of
  stable {{N{\'e}el}} skyrmions in cobalt/palladium multilayers with
  {{Lorentz}} transmission electron microscopy, Nature Communications 8 (2017)
  14761.
\newblock \href {http://dx.doi.org/10.1038/ncomms14761}
  {\path{doi:10.1038/ncomms14761}}.

\bibitem{schott_skyrmion_2017}
M.~Schott, A.~Bernand-Mantel, L.~Ranno, S.~Pizzini, J.~Vogel, H.~B{\'e}a,
  C.~Baraduc, S.~Auffret, G.~Gaudin, D.~Givord, The {{Skyrmion Switch}}:
  {{Turning Magnetic Skyrmion Bubbles}} on and off with an {{Electric Field}},
  Nano Letters 17~(5) (2017) 3006--3012.
\newblock \href {http://dx.doi.org/10.1021/acs.nanolett.7b00328}
  {\path{doi:10.1021/acs.nanolett.7b00328}}.

\bibitem{hrabec_current-induced_2017}
A.~Hrabec, J.~Sampaio, M.~Belmeguenai, I.~Gross, R.~Weil, S.~M. Ch{\'e}rif,
  A.~Stashkevich, V.~Jacques, A.~Thiaville, S.~Rohart, Current-induced skyrmion
  generation and dynamics in symmetric bilayers, Nature Communications 8 (2017)
  ncomms15765.
\newblock \href {http://dx.doi.org/10.1038/ncomms15765}
  {\path{doi:10.1038/ncomms15765}}.

\bibitem{yu_room-temperature_2016}
G.~Yu, P.~Upadhyaya, X.~Li, W.~Li, S.~K. Kim, Y.~Fan, K.~L. Wong,
  Y.~Tserkovnyak, P.~K. Amiri, K.~L. Wang, Room-{{Temperature Creation}} and
  {{Spin}}\textendash{}{{Orbit Torque Manipulation}} of {{Skyrmions}} in {{Thin
  Films}} with {{Engineered Asymmetry}}, Nano Letters 16~(3) (2016) 1981--1988.
\newblock \href {http://dx.doi.org/10.1021/acs.nanolett.5b05257}
  {\path{doi:10.1021/acs.nanolett.5b05257}}.

\bibitem{litzius_skyrmion_2017}
K.~Litzius, I.~Lemesh, B.~Kr{\"u}ger, P.~Bassirian, L.~Caretta, K.~Richter,
  F.~B{\"u}ttner, K.~Sato, O.~A. Tretiakov, J.~F{\"o}rster, R.~M. Reeve,
  M.~Weigand, I.~Bykova, H.~Stoll, G.~Sch{\"u}tz, G.~S.~D. Beach, M.~Kl{\"a}ui,
  Skyrmion {{Hall}} effect revealed by direct time-resolved {{X}}-ray
  microscopy, Nature Physics 13~(2) (2017) 170--175.
\newblock \href {http://dx.doi.org/10.1038/nphys4000}
  {\path{doi:10.1038/nphys4000}}.

\bibitem{woo_spin-orbit_2017}
S.~Woo, K.~M. Song, H.-S. Han, M.-S. Jung, M.-Y. Im, K.-S. Lee, K.~S. Song,
  P.~Fischer, J.-I. Hong, J.~W. Choi, B.-C. Min, H.~C. Koo, J.~Chang,
  Spin-orbit torque-driven skyrmion dynamics revealed by time-resolved
  {{X}}-ray microscopy, Nature Communications 8 (2017) 15573.
\newblock \href {http://dx.doi.org/10.1038/ncomms15573}
  {\path{doi:10.1038/ncomms15573}}.

\bibitem{kim_current-driven_2017-1}
J.-V. Kim, M.-W. Yoo, Current-driven skyrmion dynamics in disordered films,
  Applied Physics Letters 110~(13) (2017) 132404.
\newblock \href {http://dx.doi.org/10.1063/1.4979316}
  {\path{doi:10.1063/1.4979316}}.

\bibitem{aballe_alba_2015}
L.~Aballe, M.~Foerster, E.~Pellegrin, J.~Nicolas, S.~Ferrer, The {{ALBA}}
  spectroscopic {{LEEM}}-{{PEEM}} experimental station: Layout and performance,
  Journal of Synchrotron Radiation 22~(3) (2015) 745--752.
\newblock \href {http://dx.doi.org/10.1107/S1600577515003537}
  {\path{doi:10.1107/S1600577515003537}}.

\bibitem{foerster_custom_2016}
M.~Foerster, J.~Prat, V.~Massana, N.~Gonzalez, A.~Fontsere, B.~Molas,
  O.~Matilla, E.~Pellegrin, L.~Aballe, Custom sample environments at the {{ALBA
  XPEEM}}, Ultramicroscopy 171 (2016) 63--69.
\newblock \href {http://dx.doi.org/10.1016/j.ultramic.2016.08.016}
  {\path{doi:10.1016/j.ultramic.2016.08.016}}.

\bibitem{vansteenkiste_design_2014}
A.~Vansteenkiste, J.~Leliaert, M.~Dvornik, M.~Helsen, F.~Garcia-Sanchez, B.~V.
  Waeyenberge, The design and verification of {{MuMax3}}, AIP Advances 4~(10)
  (2014) 107133.
\newblock \href {http://dx.doi.org/10.1063/1.4899186}
  {\path{doi:10.1063/1.4899186}}.

\bibitem{rohart_path_2016}
S.~Rohart, J.~Miltat, A.~Thiaville, Path to collapse for an isolated
  {{N{\'e}el}} skyrmion, Physical Review B 93~(21) (2016) 214412.
\newblock \href {http://dx.doi.org/10.1103/PhysRevB.93.214412}
  {\path{doi:10.1103/PhysRevB.93.214412}}.

\bibitem{zhang_skyrmion-skyrmion_2015-1}
X.~Zhang, G.~P. Zhao, H.~Fangohr, J.~P. Liu, W.~X. Xia, J.~Xia, F.~J. Morvan,
  Skyrmion-skyrmion and skyrmion-edge repulsions in skyrmion-based racetrack
  memory, Scientific Reports 5 (2015) 7643.
\newblock \href {http://dx.doi.org/10.1038/srep07643}
  {\path{doi:10.1038/srep07643}}.

\bibitem{Note1}
A similar field dependence of the skyrmions size has been observed in larger
  structures, as reported by S. Woo \protect \textit {et al.} in 2 {\textmu
  }m-diameter Pt/Co/Ta disks (see \cite {woo_observation_2016-1}, Supplementary
  Information). Nevertheless, in our case, the skyrmions number and their size
  are limited by the track width at small fields.

\bibitem{ranjan_grain_1987}
R.~Ranjan, D.~C. Jiles, O.~Buck, R.~B. Thompson, Grain size measurement using
  magnetic and acoustic {{Barkhausen}} noise, Journal of Applied Physics 61~(8)
  (1987) 3199--3201.
\newblock \href {http://dx.doi.org/10.1063/1.338900}
  {\path{doi:10.1063/1.338900}}.

\bibitem{yu_pinning_1999}
R.~H. Yu, S.~Basu, Y.~Zhang, A.~Parvizi-Majidi, J.~Q. Xiao, Pinning effect of
  the grain boundaries on magnetic domain wall in {{FeCo}}-based magnetic
  alloys, Journal of Applied Physics 85~(9) (1999) 6655--6659.
\newblock \href {http://dx.doi.org/10.1063/1.370175}
  {\path{doi:10.1063/1.370175}}.

\bibitem{voto_effects_2016}
M.~Voto, L.~Lopez-Diaz, L.~Torres, Effects of grain size and disorder on domain
  wall propagation in {{CoFeB}} thin films, Journal of Physics D: Applied
  Physics 49~(18) (2016) 185001.
\newblock \href {http://dx.doi.org/10.1088/0022-3727/49/18/185001}
  {\path{doi:10.1088/0022-3727/49/18/185001}}.

\bibitem{garcia-sanchez_skyrmion-based_2016}
F.~Garcia-Sanchez, J.~Sampaio, N.~Reyren, V.~Cros, J.-V. Kim, A skyrmion-based
  spin-torque nano-oscillator, New Journal of Physics 18~(7) (2016) 075011.
\newblock \href {http://dx.doi.org/10.1088/1367-2630/18/7/075011}
  {\path{doi:10.1088/1367-2630/18/7/075011}}.

\bibitem{pulecio_hedgehog_2016}
J.~F. Pulecio, A.~Hrabec, K.~Zeissler, R.~M. White, Y.~Zhu, C.~H. Marrows,
  Hedgehog {{Skyrmion Bubbles}} in {{Ultrathin Films}} with {{Interfacial
  Dzyaloshinskii}}-{{Moriya Interactions}}, arXiv:1611.06869 [cond-mat]\href
  {http://arxiv.org/abs/1611.06869} {\path{arXiv:1611.06869}}.

\end{thebibliography}

\end{document}